# Achieving 100% amplitude modulation depth in a graphene-based tuneable capacitance metamaterial

Ruqiao Xia[a,1], Nikita W. Almond[a], Stephen J. Kindness[a], Sergey A. Mikhailov[b], Wadood Tadbier[c], Riccardo Degl'Innocenti[d], Yuezhen Lu[e], Abbie Lowe[a], Ben Ramsay[a], Lukas A. Jakob[a], James Dann[a], Stephan Hofmann[c], Harvey E. Beere[a], David A. Ritchie[a,f], Wladislaw Michailow[a,2]

[a:] Cavendish Laboratory, University of Cambridge, CB3 0HE Cambridge, UK

[b:] Institute of Physics, University of Augsburg, D-86135 Augsburg, Germany

[c:] Department of Engineering, University of Cambridge, Cambridge, UK

[d:] School of Electronic Engineering and Computer Science, Queen Mary University of London, Mile End Road, London E1 4NS, UK

[e:] Department of Electronic & Electrical Engineering, Faculty of Engineering Sciences, University College London, London, WC1E 7JE, UK

[f:] Swansea University, Singleton Park, Sketty, Swansea SA2 8PP, UK

Electronic mails: 1 - rx224@cam.ac.uk ; 2 - wm297@cam.ac.uk

## Abstract

Effective control of terahertz radiation requires the development of efficient and fast modulators with a large modulation depth. This challenge is often tackled by using metamaterials, artificial sub-wavelength optical structures engineered to resonate at the desired terahertz frequency. Metamaterial-based devices exploiting graphene as the active tuneable element have been proven to be a highly effective solution for THz modulation. However, whilst the graphene conductivity can be tuned over a wide range, it cannot be reduced to zero due to the gapless nature of graphene, which directly limits the maximum achievable modulation depth for single-layer metamaterial modulators. Here, we demonstrate two novel solutions to circumvent this restriction: Firstly, we excite the modulator from the back of the substrate, and secondly, we incorporate air gaps into the graphene patches. This results in a ground-breaking graphene-metal metamaterial terahertz modulator, operating at 2.0–2.5 THz, which demonstrates a 99.01 % amplitude and a 99.99 % intensity modulation depth at 2.15 THz, with a reconfiguration speed in excess of 3 MHz. Our results open up new frontiers in the area of terahertz technology.

**Keywords:** Graphene, 100% modulation depth, metamaterial, metasurface, back excitation, air gap, reflection, terahertz modulator, high speed modulator.

# Introduction

Terahertz (THz) technology has a myriad of applications in areas including medicine [1, 2], security [3–5], and wireless networks [6–8], due to the unique advantages that THz waves can offer in spectroscopy, non-destructive testing, imaging, and communications. These applications require electrically tuneable modulators to manipulate the specific properties of electromagnetic waves, such as the amplitude, phase, or polarization of the electric field.

One way to realise such modulators is to use electrically tuneable metamaterials, which consist of artificially engineered subwavelength structures with electrically tuneable elements [9–11]. These elements allow the optical properties of the metamaterials to be dynamically adjusted [12–20].

In the pursuit of effective THz modulation, the two-dimensional material graphene stands out as a promising candidate for attaining the desired tunability, thanks to its large, electrically controllable conductivity [21–25]. The current commercial availability of large-scale CVD-grown graphene [26–29] enables its integration into devices in a scalable and CMOS-compatible fabrication process [30–32]. Significant efforts have been put into achieving high modulation depth through various methodologies [33–39].

However, graphene has one peculiarity which poses a problem in achieving a large modulation depth: even at the Dirac point, a conducting graphene channel cannot be fully depleted. This results in a lower limit of the conductivity, which directly translates into a restriction on the modulation depth of single-layer THz modulators. To overcome this inherent limitation, various approaches have been proposed. For external graphene-loaded modulators, i.e. devices not integrated with a source, 90% intensity modulation depth has been achieved by enhancing the metamaterial response through a quarter-wave resonance cavity, with a bandwidth of 20 kHz [40, 41]. Broadband THz modulators based on graphene/ionic liquid/graphene sandwich structures were demonstrated with 83% and 93% intensity modulation depth for single-layer and trilayer graphene-based devices, respectively [42]. Another example involves a broadband graphene THz modulator that exploits graphene to control the Brewster angle [43]. This modulator demonstrates an intensity modulation depth of 99.3-99.9 % with 6 kHz modulation speed within a few degrees of the Brewster angle. While the above-mentioned examples show promising modulation performance with speeds up to a few tens of kHz, the slow reconfiguration speeds of ionic liquid gating and the large capacitance resulting from the large area graphene pose intrinsic limitations on the maximum possible speeds. However, in order for a modulator to be practically useful in high-speed applications, such as real-time imaging and communication, the modulation must rely on an intrinsically fast mechanism, with response times in the MHz-GHz range. In the area of high-speed modulation, 100 % modulation depth with a 100 MHz speed has been demonstrated, but in a graphene modulator integrated with a THz quantum cascade laser operating at cryogenic temperatures [44]. At sub-THz frequencies (0.35 THz), gallium nitride-based modulators have been reported with speeds up to 1 – 3 GHz [45, 46]. Recently, graphene-based metamaterial modulators operating at 0.75 THz have shown promise for several GHz modulation speeds with a modulation depth of 7.6 dB [47]. Thus, the search for a fast, room-temperature standalone THz modulator capable of achieving 100% amplitude modulation depth is still ongoing.

In this work, we demonstrate two novel improvements which enable 100% amplitude modulation to be realised. Firstly, we use "back excitation" of the modulator, by exciting it from the rear substrate side. This allows us to exploit destructive interference of Fresnel reflection components from the substrate and the metamaterial. Secondly, instead of shunting the metamaterial gaps with graphene and exploiting its resistive dampening, we design our structure to utilize graphene as a tuneable capacitor by incorporating air gaps in the graphene patches. We demonstrate these solutions in a single-layer, all-solid-state, graphene-metal metamaterial modulator operating at a central frequency of 2.15THz at room temperature. The modulator exhibits 99.01 % amplitude and 99.99 % intensity modulation depth, corresponding to 40.1 dB, with reconfiguration speeds in excess of 3 MHz.

## Results

### Device structure and operating principle

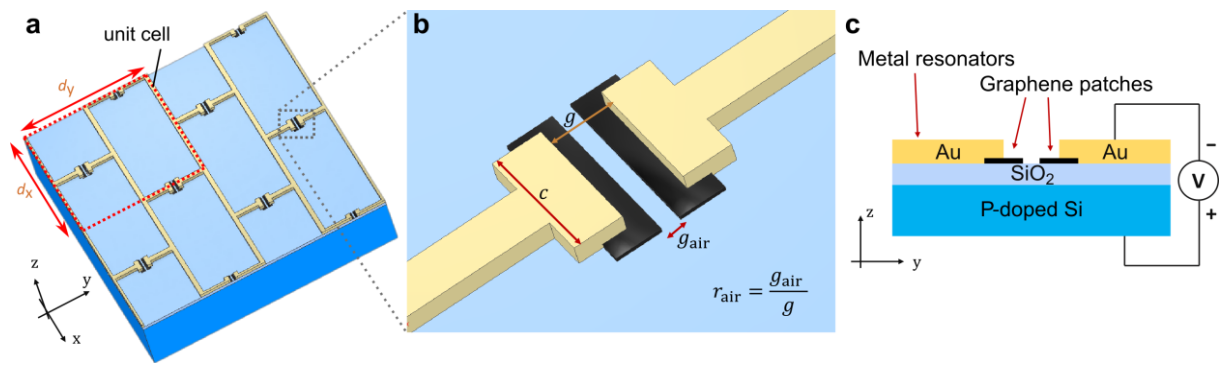

**Figure 1: Schematic of the brickwork antenna tuneable capacitance modulator.** (a) 3D view of the device area, the unit cell is indicated by a red dashed box, (b) an enlarged view of a metamaterial capacitor with the air gap in the graphene patch, and (c) a cross-sectional view of the device with the back-gate voltage source indicated.

The metamaterial design is shown in Fig. 1 (a). The structure is designed for exposure to normally incident, y-polarized THz radiation. It comprises a brickwork antenna array with capacitive gaps built into the antenna elements parallel to the y-direction. The incident radiation generates an induced current in the structure, and the metal antennas confine the electric field (E-field) in the resonator gaps. Graphene patches are placed within the gaps and serve as the active elements, see Fig. 1 (b). The dimensions of the unit cell are $d_x$ and $d_y$, the capacitor width and gap length are *c* and *g*, respectively. The length of the air gap in the graphene is $g_{air}$. The conductivity of the graphene patches can be continuously tuned by applying a bias voltage between the back of the p-doped silicon substrate and the metal resonator array, see Fig. 1 (c).

The modulation depth is one of the most critical figures of merit that must be considered when designing and evaluating a modulator. The modulation depth $h$ is quantified by the expression:

$$h = \frac{A_{\text{Max}} - A_{\text{Min}}}{A_{\text{Max}}}, \quad (1)$$

where $A_{\text{Max}}$ and $A_{\text{Min}}$ are the maximum and minimum values of an optical parameter that can be electrically controlled. This parameter, for example, could be the electric field amplitude of the reflected or transmitted wave. 100% modulation depth can be realized when the transmission or reflection vanishes at resonance; in this case $A_{\text{Min}} = 0$.

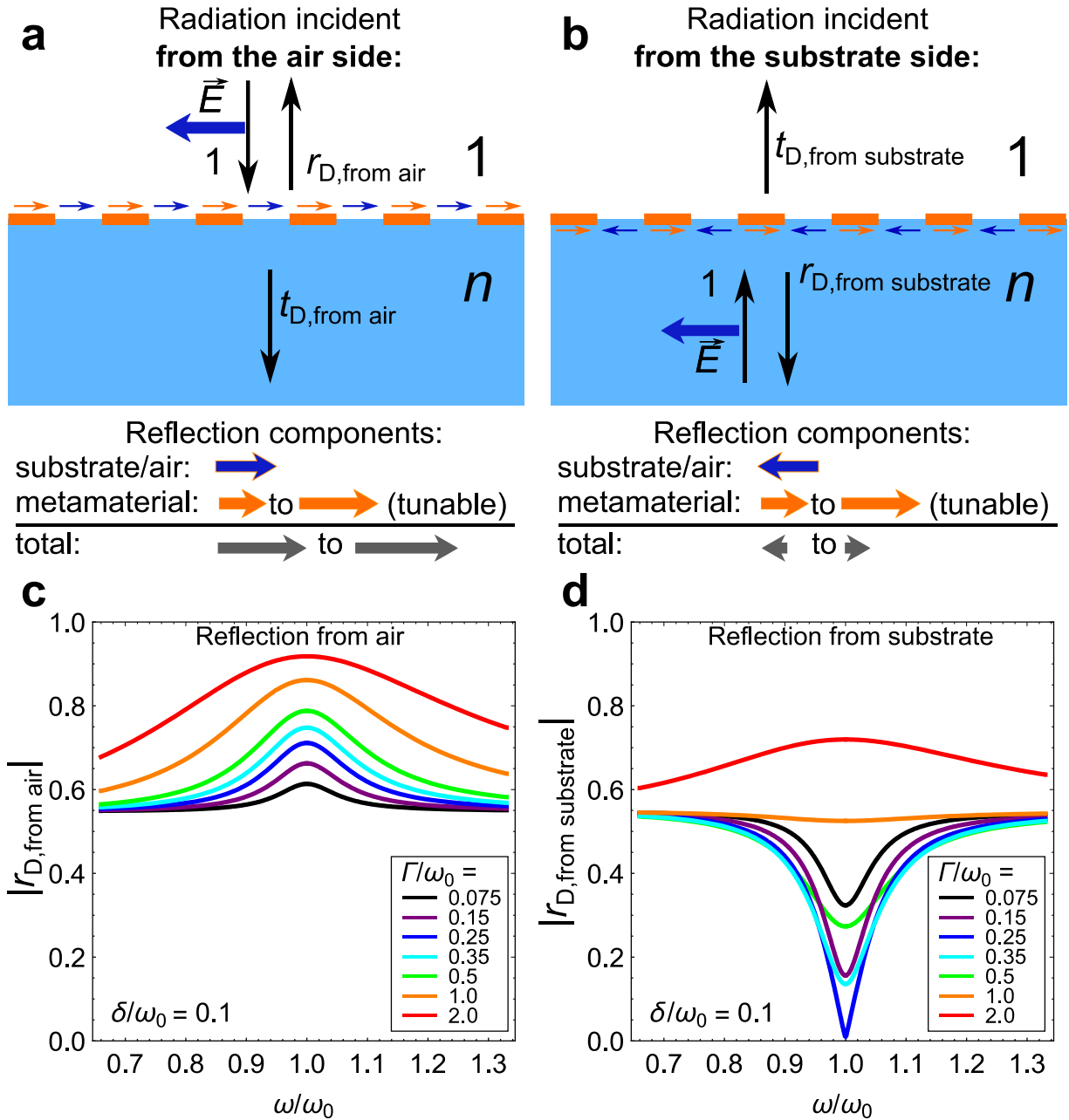

**Figure 2: Reflection from a metamaterial on a substrate with refractive index *n*.** Comparison between the excitation (a), (c) from the top-side and (b), (d) from the back-side. (a), (b): Schematic diagram of the metamaterial reflection. Vertical arrows show the direction of the electromagnetic wave vector; the horizontal arrows indicate the electric field direction. The coloured arrows indicate the direction of the reflected electric field from the substrate-air interface (blue) and from the metamaterial (orange). (c), (d): The absolute value of the analytically calculated reflection coefficient, according to the model in Eqs. (3), (4), as a function of the normalised frequency $\omega/\omega_0$ for (c) top and (d) back excitation, for $\delta/\omega_0$ = 0.1 and different values of $\Gamma/\omega_0$.

The vast majority of graphene modulators rely on top-side excitation, with the wave incident on the device from the metamaterial side. In this case, the reflection or transmission cannot be fully extinguished; however, in the case of the back-excitation reflection, it is possible to turn $A_{\text{Min}}$ to zero. This can be shown as follows. The reflection from a metamaterial described by an effective 2D conductivity $\sigma(\omega)$ on a substrate with refractive index *n* can be calculated using Maxwell's equations. The transmission coefficient is the same regardless of whether the light is incident from the substrate side or from the air side,

$$t_{\text{D,from air}} = t_{\text{D,from substrate}} = \frac{2}{n+1+S}. \quad (2)$$

Here, $S = c\mu_0\sigma(\omega)$ is a dimensionless value proportional to the effective conductivity of the metamaterial. But the reflection coefficients are different: in the case of incidence from the air side it is

$$r_{\text{D,from air}} = \frac{-1}{n+1+S}[(n-1)+S], \quad (3)$$

and for radiation incident from the substrate side, it is:

$$r_{\text{D,from substrate}} = \frac{1}{n+1+S}[(n-1)-S]. \quad (4)$$

As can be seen, the transmission can be minimized by maximising *S*, but will never fully reach zero. The reflection coefficients consist of two terms: one with (*n*-1), corresponding to the Fresnel reflection due to the substrate-air refractive index mismatch, and another one with *S*, representing the reflection from the metamaterial. Eq. (3) contains their sum, while Eq. (4) contains their difference. This means that under certain conditions, the metamaterial reflection from the substrate side can vanish, resulting in $A_{\text{min}}$ = 0.

This phenomenon is illustrated in Fig. 2 (a), (b). The electric fields of the waves reflected from the metamaterial are shown by the orange arrows, and those from the Fresnel reflection due to the substrate-air refractive index mismatch are shown by the blue arrows. The total reflection at the interface between air and the metamaterial/substrate is determined by the vector sum of these two components. Hence, in the case where light is incident from the air side, the Fresnel air-substrate and the metamaterial reflection components interfere constructively, Fig. 2 (a). Therefore, any change in the optical properties due to the metamaterial will result in only a weak change in the total reflection amplitude. In contrast, if the light is incident from the substrate side, the Fresnel components from the metamaterial and the air-substrate contributions interfere destructively, Fig. 2 (b). In this case, any small change in metamaterial optical properties will be greatly enhanced, enabling 100% amplitude modulation depth.

The solution of electrodynamic equations describing the resonance behaviour of the metamaterial structure results in the frequency dependencies of the absolute values of the reflection coefficients obtained from Eqs. (3) and (4) in Figs. 2(c) and (d). The resonance is at the frequency $\omega = \omega_0$, the peak width is proportional to $\delta$, and $\Gamma$ is a factor related to the oscillator strength. Only when the device is excited from the substrate-side, can an extinction of the reflection be observed at the resonant frequency at a certain value of $\Gamma/\delta$.

A second crucial improvement that we introduce is an air gap in the graphene patch in the centre of the capacitor, see Fig. 1(b), showing the graphene patches protruding from either side of the capacitor. A metamaterial can be understood as a planar realisation of LC-type resonators. If graphene is used to shunt the capacitor gaps in the resonators, it is used as a tuneable resistor. The resonator gap corresponds to a capacitor with variable loss in this case, and the tuning of the graphene conductivity results in a broadening of the resonance. By introducing an air gap in the graphene patch, the graphene is instead made to serve as a tuneable capacitor. This way, as we will see in the following, the resonance shifts while maintaining its Q-factor. Not only does this allow for a higher modulation depth, but also provides frequency tuneability.

The unit cell of the brickwork antenna metamaterial structure, shown in Fig. 1 (a), is simulated using the RF module of COMSOL Multiphysics software (version 6.1). The antenna parameters have been tailored to produce a resonant frequency at 2.4 THz without graphene present in the gaps. The resulting parameters are $g$ = 1.5 µm, $c$ = 3 µm, $d_\mathrm{x}$ = 35 µm, $d_\mathrm{y}$ = 35 µm. The width of all metallic lines in the x-y-plane is 1.2 µm, and we choose an exemplary air gap ratio of $r_\mathrm{air} = g_\mathrm{air}/g = 0.3$. The Drude model is used to simulate the behaviours of the conductive materials, with the parameters from Ref. [48] (graphene) and Refs. [49, 50] (gold).

The simulated reflection amplitudes from the brickwork antenna metamaterial structure for waves incident from the substrate against frequency at different conductivities is shown in Fig. 3 (a). When the active tuneable element has zero conductivity (corresponding to a pure metallic brickwork antenna without graphene), the reflected electric field reaches its maximum value at resonance. As the conductivity of the graphene increases, the maximum of the reflection curve at resonance is converted to a minimum in the range from 0.01 mS to 0.1 mS. As the conductivity reaches values around 0.3 mS to 0.5 mS, the reflection vanishes at the resonance. This pattern is repeated in reverse order as the graphene conductivity keeps increasing. The higher the graphene conductivity becomes above ca. 4 mS, the more the reflection approaches the shape expected from a pure metallic resonance with a reduced capacitive gap. The general trend seen from simulations, Fig. 3(a), is in agreement with the analytical curves, Fig. 2(d). There is, however, a difference: the resonance frequency shifts as the graphene conductivity is varied, as the increase in conductivity of the active element corresponds to an effective reduction of the capacitor gap size. At this point, the active element shifts the resonant frequency with no damping. This phenomenon, not considered in our analytical model in Eq. (4), arises due to the air gaps.

Let us now consider the difference between the case when the graphene shunts the capacitors without an air gap and the case when an air gap is introduced in the graphene patch. Figure 3(b) compares the absolute values of the reflection coefficients of the brickwork antenna structure as a function of graphene conductivity for three different cases, top-excitation and no air gap (blue), back-excitation and no air gap (black), and back-excitation with an air gap (red). The reflection stays above the Fresnel substrate-air reflection value in the top-excitation case. In the back-excitation case, the reflection vanishes both with and without an air gap. The slope is more severe when there is an air gap present, allowing a higher modulation depth to be achieved for a given graphene conductivity. It is also possible to achieve a higher reflection with an air gap, which reduces the insertion loss of the device.

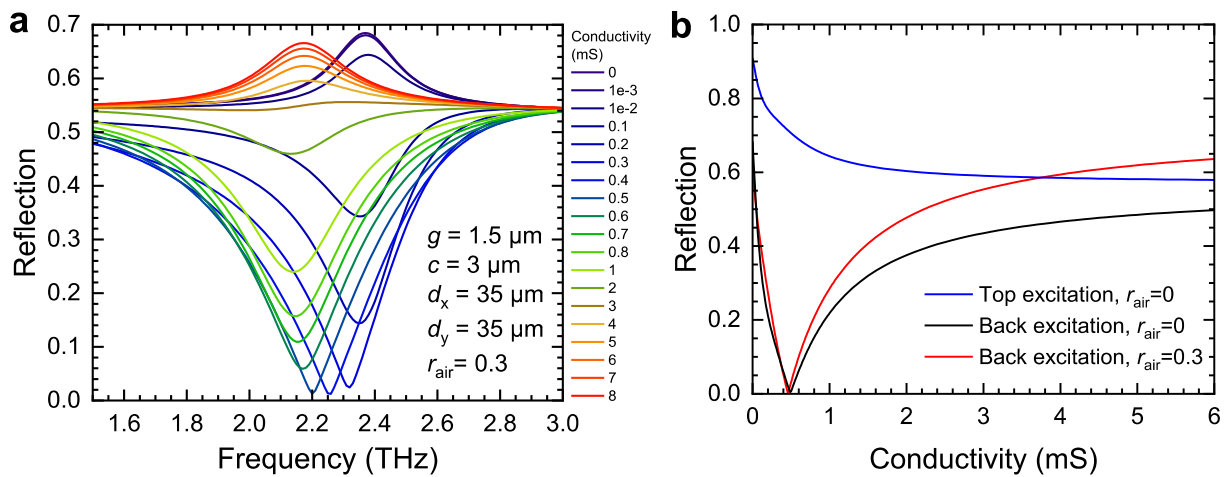

**Figure 3: Numerical simulations of the reflected electric field amplitudes.** The simulated structure has the parameters $g$ = 1.5 μm, $c$ = 3 μm, $d_x$ = 35 μm, $d_y$ = 35 μm. (a) Reflection against frequency for an air gap ratio of $r_{\text{air}} = 0.3$. (b) Reflection against graphene conductivity. For the case $r_{\text{air}} = 0$, i.e. no air gap, the blue and black curves show the reflection at the resonant frequency of the pure metallic brickwork antenna structure of 2.38 THz for the top and back excitation cases. The red curve

corresponds to $r_\text{air} = 0.3$ and the frequency 2.22 THz, where the minimum of the back excitation reflection is observed.

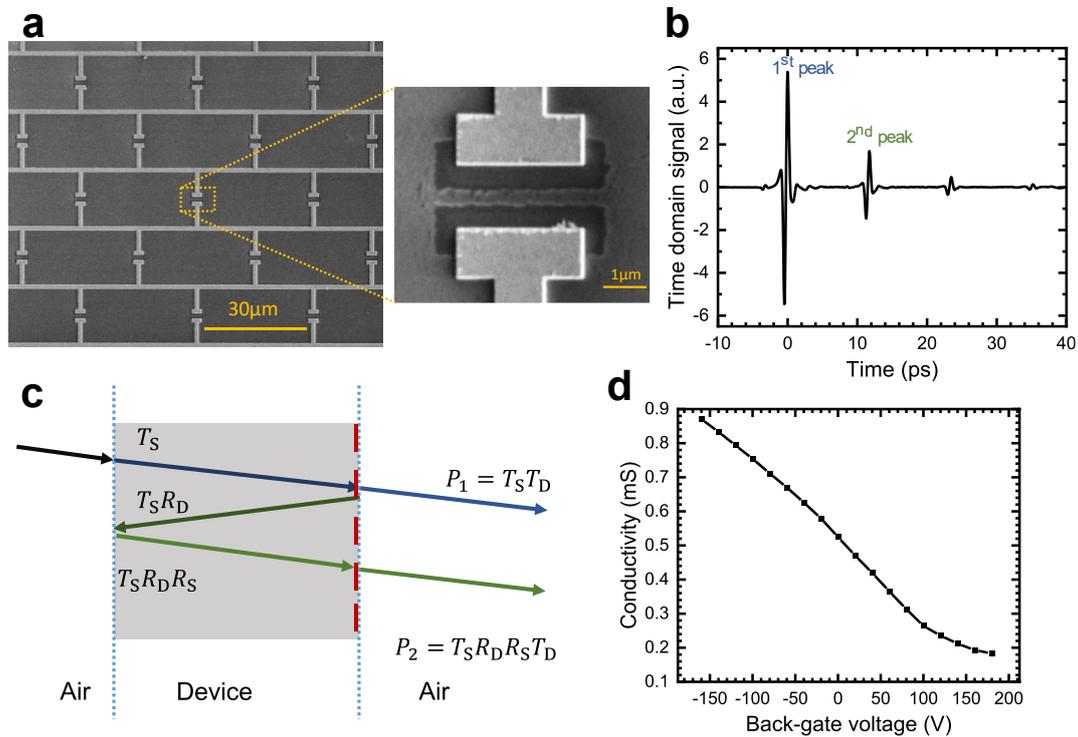

**Figure 4: Characterisation and measurement of the fabricated tuneable capacitance modulator.** (a) Scanning electron microscope (SEM) image of the brickwork antenna arrays, with a zoomed-in view of the capacitor and air gap. (b) Measured transmitted pulses in the time domain, showing the different peaks resulting from internal reflections within the device. (c) Schematic diagram of the internal reflections inside the substrate and the corresponding signals, where $T_\text{D}$ and $R_\text{D}$ are the transmittance and reflectance of the metamaterial, and $T_\text{S}$ and $R_\text{S}$ are the transmittance and reflectance of the silicon substrate. The device is excited from the substrate side under normal incidence, and the metamaterial is indicated by the dashed red line. (d) Measured graphene conductivity as a function of the back-gate voltage applied across substrate and metal contacts.

### Fabrication and measurement

We fabricate the tuneable capacitance modulator with a brickwork antenna structure and graphene patches with air gap. First, the graphene is transferred using a wet transfer technique onto a 550 μm boron p-doped high resistive silicon substrate (>100 Ωcm) with a 300 nm SiO₂ layer on top. The graphene used in this device is monolayer graphene on copper, grown by chemical vapour deposition, procured from a commercial supplier (Graphenea). The graphene patches are created using electron beam lithography and subsequent selective etching of the graphene in an oxygen plasma. The titanium-gold metal layer of the brickwork arrays (14nm Ti, 156nm Au) is thermally evaporated after a second electron beam lithography onto the sample with an active device area of 1.3 mm by 1.3 mm, comprising 36 × 36 unit cells. A scanning electron microscope (SEM) image of the fabricated device is shown in Fig. 4 (a). The device has an air gap ratio of approximately $r_\text{air} = 0.23$, as measured from SEM images; the other design parameters are displayed in Fig. 3 (a).

To measure the reflectance of the metamaterial at the excitation from the substrate side, we use a terahertz time-domain spectroscopy (TDS) setup, Tera K15 from Menlo Systems, at room temperature in a dehumidified dry air environment (150 ppm by volume, at the same impurity level as commercial gaseous nitrogen). The TDS measurement of the transmitted electric field shows several pulses in the time domain, see Fig 4(b). The first and second peaks are selected using a Blackman-Harris window function with a 12 ps full width. The reflectance is extracted from the ratio of the squared absolute values of the Fourier transforms of the second and first transmitted peaks, $P_2$ and $P_1$. As seen from Fig. 4(c), their ratio equals

$$\frac{P_2}{P_1} = \frac{T_\text{D}\,T_\text{S} R_\text{S} R_\text{D}}{T_\text{D}\,T_\text{S}} = R_\text{S} R_\text{D}, \qquad (5)$$

where $R_\text{D}$ is the reflectance when the device is excited from the back side (substrate side), and $R_\text{S}$ is the reflectance of the substrate and air interface. To obtain the reflectance of the device $R_\text{D}$, this value is divided by the reflectance of the silicon substrate $R_\text{S} = [(n\text{-}1)/(n\text{+}1)]^2 = 0.300$, with $n$ = 3.425 [51–53].

The obtained reflectance $R_\text{D}$ is shown in Fig. 5(a) for different back-gate voltages. The device achieves a modulation depth as high as 99.01 % in amplitude and 99.99 % in intensity, corresponding to 40.1 dB, at a frequency of 2.15 THz.

The experimental data, shown on a logarithmic scale in Fig. 5(a), reveals a fine structure of the minimum, which is split in two. Fig. 5(b) shows the results of simulations performed for the fabricated device with the air gap ratio of $r_\text{air} = 0.23$ and the measured graphene conductivity range of 0.18 mS to 0.90 mS. The conductivity is estimated using two-terminal measurements on a 70 µm by 200 µm graphene test pad close to the modulator structure, but on the same chip, having undergone the same cleanroom fabrication process, and is shown in Fig. 4 (d) as a function of the back-gate voltage. This data was recorded simultaneously with the TDS measurement.

There is a remarkably good agreement between experimentally measured reflectance and the numerically simulated data. Both the overall trend, as well as the expected modulation ranges, are well reproduced. The turning point, where the reflectance is fully extinguished (less than 1% in amplitude), is contained within the attainable graphene conductivity range of 0.18 mS to 0.90 mS. It should be noted that the plotted conductivities in the simulation cannot be expected to match exactly to the measured values on the 70 μm × 200 μm graphene test area, due to effect such as the contact resistance at the metal-graphene interface [54].

To ensure reproducibility of the results, we repeated the measurements 6 times, including measurements on different days and with re-alignment of the sample position. The device has consistently demonstrated 97.65 – 99.41 % amplitude and 99.94 – 100 % intensity modulation depth as the back-gate voltage was changed from -160 V to 180 V.

The frequency dependence of the measured modulation depth and insertion loss is shown in Fig. 5 (c). The tuning ranges are extracted for each frequency point. Due to the tuneable resonance effect originating from the air gap, the maximum insertion loss and the maximum modulation depth are achieved at different frequencies.

The speed of the metamaterial-based modulators is limited by the RC time constant of the structure and can be measured by applying an AC sine wave of varying frequency to the device. As the frequency of the driving signal grows, the change between the data with and without AC signal will diminish in an RC-type roll-off above the -3 dB point [55,16]. To measure the modulation speed of the device, we use a TG320 function generator to apply a 21.4 V peak-to-peak AC signal to the back gate at zero DC bias. The transmitted E-fields are measured using TDS with and without the AC signal. The difference is evaluated by integrating the ratio of the transmission with and without AC voltage applied from 1.9 THz to 2.1 THz, where the maximum difference between the two cases is observed, see Fig. 5 (d). The error bars represent the standard deviation from 6 measurements. No roll-off can be observed up to 3 MHz, which is the maximum frequency of the TG320 function generator. This shows that the modulation speed of the device exceeds 3 MHz.

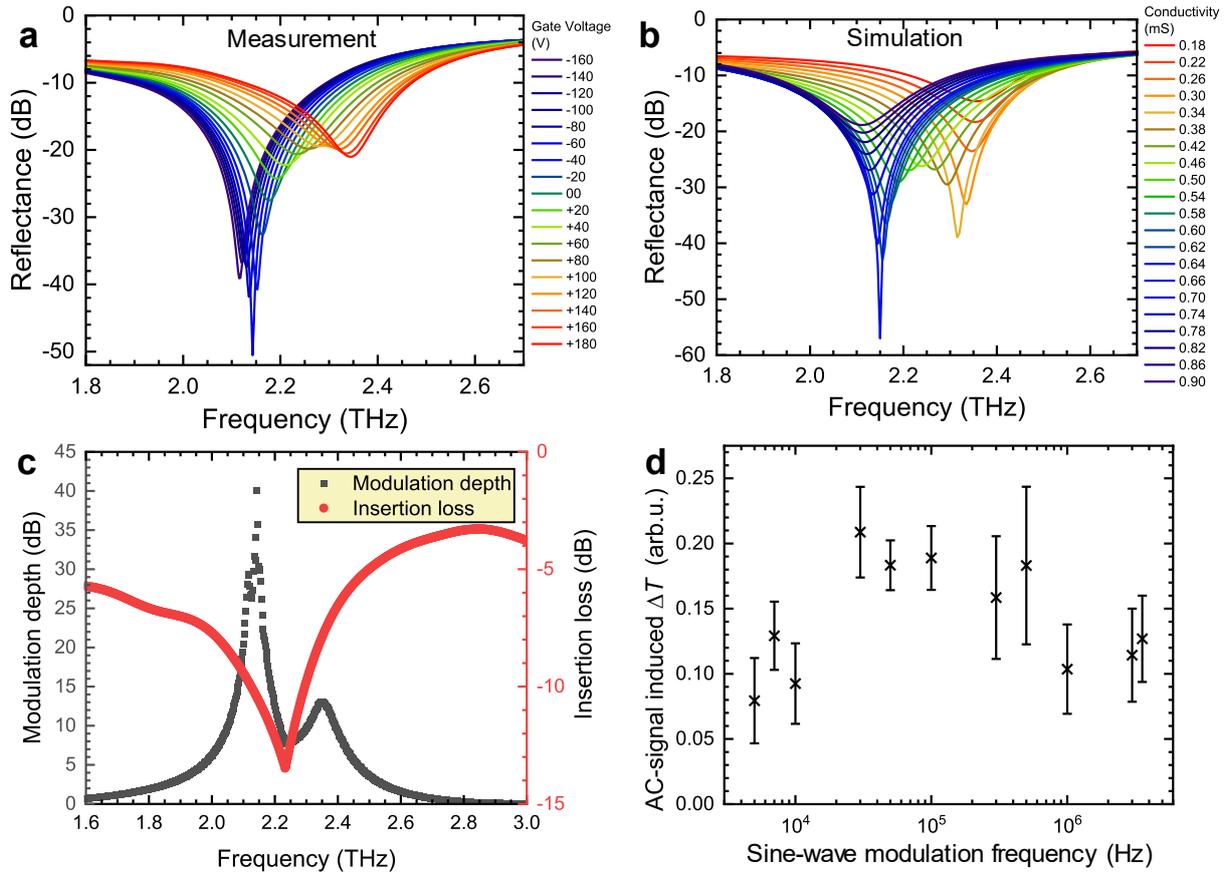

**Figure 5: Experimental results and comparison with simulation.** (a), (b) Reflectance of the brickwork antenna tuneable capacitance metamaterial as a function of frequency: (a) measurement with gate voltage varied from -160V to 180V, (b) numerical simulations of the structure with an air ratio gap of $r_{\text{air}} = 0.23$ for different graphene conductivities. (c) THz frequency dependence of the modulation depth and insertion loss. (d) Modulation speed evaluated from the difference between the DC and AC measurements taken by TDS, as a function of the driving AC sine wave frequency.

## Discussion and conclusion

One of the major challenges in achieving full modulation in graphene-based metamaterial modulators is the finite conductivity at the Dirac point. In response to this challenge, we proposed and experimentally demonstrated a novel solution that overcomes this limitation using excitation of the modulator from the substrate side. We further reduced the insertion losses by incorporating air gaps into the graphene patches, leading to the device acting as a tuneable capacitance modulator. As a result, we demonstrated a graphene-metal metamaterial terahertz modulator, operating within the 2–2.5 THz range, which achieves 99.01 % amplitude and 99.99 % intensity modulation depth at 2.15 THz with reconfiguration speeds in excess of 3 MHz. To the best of our knowledge, this is the highest reported modulation depth for a graphene-based metamaterial that is not integrated with a radiation source.

The back excitation approach to achieve full modulation is universal and is not limited to graphene-based devices. It can be employed in a multitude of metamaterial modulator device architectures, employing other semiconductors as the active element, and can also be used to realise phase modulation with a large depth [56].

The introduction of an air gap also led to a resonant frequency shift. This phenomenon holds promise for diverse applications, such as metamaterial-based tuneable bandpass or notch filters. Employing capacitive tuning instead of resistive shorting also has the potential to boost the modulation depth in transmission, both in graphene-based and in other, semiconducting, 2DEG channels. Moreover, the demonstrated gigahertz modulation speed of single-layer metamaterial modulators [45,46,47] underscores this device's potential for high-speed modulation applications. Our approach does not rely on external cavities for modulation enhancement and employs metamaterials that have the potential to operate at GHz speeds.

## Acknowledgements

The authors thank David Ellis for wafer cutting, Jonathan Griffiths for help with electron beam lithography, and Richard Langford and Eric Tapley for help with scanning electron microscopy. W.M. thanks Trinity College Cambridge for a Junior Research Fellowship. The authors acknowledge EPSRC funding from the HyperTerahertz grant, no. EP/P021859/1, and the TeraCom grant, no. EP/W028921/1.

## Author contributions

R.X. carried out the numerical simulations, fabricated and measured the devices, analysed and visualised the data. N.A. set up the experimental system, advised, provided numerical analysis software, and edited the manuscript. R.X. and W.M. designed the devices, interpreted the data, and wrote the manuscript. W.T. and S.H. advised on transfer of CVD-grown graphene-on-copper. J.D. did electron beam lithography exposure, Y.L. did reactive ion etching of silicon dioxide, and A.L. and B.R. did hydrogen fluoride etching during sample fabrication. L.J. wrote software for graphene pre-characterisation. S.M. provided theoretical support. R.D. and H.E.B. advised and commented on the manuscript. D.A.R. acquired funding, provided resources, and managed the grants. W.M. and S.J.K. conceived the device concept. W.M. supervised the project. All authors discussed the results and the manuscript.